\documentclass[aps,prd,reprint,superscriptaddress,showpacs,showkeys]{revtex4-1}

\usepackage{graphicx}
\usepackage{natbib}
\usepackage{color}
\usepackage{subfigure}
\usepackage{enumerate}

\begin{document}


\title{Constraints on Axion couplings from the CDEX-1 experiment at the China Jinping Underground Laboratory}


\affiliation{Key Laboratory of Particle and Radiation Imaging (Ministry of Education) and Department of Engineering Physics, Tsinghua University, Beijing 100084}
\affiliation{College of Physical Science and Technology, Sichuan University, Chengdu 610064}
\affiliation{Department of Nuclear Physics, China Institute of Atomic Energy, Beijing 102413}
\affiliation{School of Physics, Nankai University, Tianjin 300071}
\affiliation{NUCTECH Company, Beijing 100084}
\affiliation{YaLong River Hydropower Development Company, Chengdu 610051}
\affiliation{Institute of Physics, Academia Sinica, Taipei 11529}
\affiliation{Department of Physics, Banaras Hindu University, Varanasi 221005}

\author{S.K. Liu}
\affiliation{Key Laboratory of Particle and Radiation Imaging (Ministry of Education) and Department of Engineering Physics, Tsinghua University, Beijing 100084}
\affiliation{College of Physical Science and Technology, Sichuan University, Chengdu 610064}
\author{Q. Yue}
\email{Corresponding author: yueq@mail.tsinghua.edu.cn}
\affiliation{Key Laboratory of Particle and Radiation Imaging (Ministry of Education) and Department of Engineering Physics, Tsinghua University, Beijing 100084}
\author{K.J. Kang}
\affiliation{Key Laboratory of Particle and Radiation Imaging (Ministry of Education) and Department of Engineering Physics, Tsinghua University, Beijing 100084}
\author{J.P. Cheng}
\affiliation{Key Laboratory of Particle and Radiation Imaging (Ministry of Education) and Department of Engineering Physics, Tsinghua University, Beijing 100084}
\author{H.T. Wong}
\altaffiliation{Participating as a member of TEXONO Collaboration}
\affiliation{Institute of Physics, Academia Sinica, Taipei 11529}
\author{Y.J. Li}
\affiliation{Key Laboratory of Particle and Radiation Imaging (Ministry of Education) and Department of Engineering Physics, Tsinghua University, Beijing 100084}
\author{H.B. Li}
\altaffiliation{Participating as a member of TEXONO Collaboration}
\affiliation{Institute of Physics, Academia Sinica, Taipei 11529}
\author{S.T. Lin}
\affiliation{College of Physical Science and Technology, Sichuan University, Chengdu 610064}
\affiliation{Institute of Physics, Academia Sinica, Taipei 11529}
\author{J.P. Chang}
\affiliation{NUCTECH Company, Beijing 100084}
\author{J.H.~Chen}
\altaffiliation{Participating as a member of TEXONO Collaboration}
\affiliation{Institute of Physics, Academia Sinica, Taipei 11529}
\author{N.~Chen}
\affiliation{Key Laboratory of Particle and Radiation Imaging (Ministry of Education) and Department of Engineering Physics, Tsinghua University, Beijing 100084}
\author{Q.H.~Chen}
\affiliation{Key Laboratory of Particle and Radiation Imaging (Ministry of Education) and Department of Engineering Physics, Tsinghua University, Beijing 100084}
\author{Y.H. Chen}
\affiliation{YaLong River Hydropower Development Company, Chengdu 610051}
\author{Z. Deng}
\affiliation{Key Laboratory of Particle and Radiation Imaging (Ministry of Education) and Department of Engineering Physics, Tsinghua University, Beijing 100084}
\author{Q. Du}
\affiliation{College of Physical Science and Technology, Sichuan University, Chengdu 610064}
\author{H. Gong}
\affiliation{Key Laboratory of Particle and Radiation Imaging (Ministry of Education) and Department of Engineering Physics, Tsinghua University, Beijing 100084}
\author{H.J. He}
\affiliation{Key Laboratory of Particle and Radiation Imaging (Ministry of Education) and Department of Engineering Physics, Tsinghua University, Beijing 100084}
\author{Q.J.~He}
\affiliation{Key Laboratory of Particle and Radiation Imaging (Ministry of Education) and Department of Engineering Physics, Tsinghua University, Beijing 100084}
\author{H.X.~Huang}
\affiliation{Department of Nuclear Physics, China Institute of Atomic Energy, Beijing 102413}
\author{H. Jiang}
\affiliation{Key Laboratory of Particle and Radiation Imaging (Ministry of Education) and Department of Engineering Physics, Tsinghua University, Beijing 100084}
\author{J.M. Li}
\affiliation{Key Laboratory of Particle and Radiation Imaging (Ministry of Education) and Department of Engineering Physics, Tsinghua University, Beijing 100084}
\author{J. Li}
\affiliation{Key Laboratory of Particle and Radiation Imaging (Ministry of Education) and Department of Engineering Physics, Tsinghua University, Beijing 100084}
\author{J. Li}
\affiliation{NUCTECH Company, Beijing 100084}
\author{X. Li}
\affiliation{Department of Nuclear Physics, China Institute of Atomic Energy, Beijing 102413}
\author{X.Q. Li}
\affiliation{School of Physics, Nankai University, Tianjin 300071}
\author{X.Y.~Li}
\affiliation{School of Physics, Nankai University, Tianjin 300071}
\author{Y.L.~Li}
\affiliation{Key Laboratory of Particle and Radiation Imaging (Ministry of Education) and Department of Engineering Physics, Tsinghua University, Beijing 100084}
\author{F.K. Lin}
\altaffiliation{Participating as a member of TEXONO Collaboration}
\affiliation{Institute of Physics, Academia Sinica, Taipei 11529}
\author{L.C.~L\"{u}}
\affiliation{Key Laboratory of Particle and Radiation Imaging (Ministry of Education) and Department of Engineering Physics, Tsinghua University, Beijing 100084}
\author{H. Ma}
\affiliation{Key Laboratory of Particle and Radiation Imaging (Ministry of Education) and Department of Engineering Physics, Tsinghua University, Beijing 100084}
\author{J.L. Ma}
\affiliation{Key Laboratory of Particle and Radiation Imaging (Ministry of Education) and Department of Engineering Physics, Tsinghua University, Beijing 100084}
\author{S.J. Mao}
\affiliation{NUCTECH Company, Beijing 100084}
\author{J.Q. Qin}
\affiliation{Key Laboratory of Particle and Radiation Imaging (Ministry of Education) and Department of Engineering Physics, Tsinghua University, Beijing 100084}
\author{J. Ren}
\affiliation{Department of Nuclear Physics, China Institute of Atomic Energy, Beijing 102413}
\author{J. Ren}
\affiliation{Key Laboratory of Particle and Radiation Imaging (Ministry of Education) and Department of Engineering Physics, Tsinghua University, Beijing 100084}
\author{X.C.~Ruan}
\affiliation{Department of Nuclear Physics, China Institute of Atomic Energy, Beijing 102413}
\author{V. Sharma}
\altaffiliation{Participating as a member of TEXONO Collaboration}
\affiliation{Institute of Physics, Academia Sinica, Taipei 11529}
\affiliation{Department of Physics, Banaras Hindu University, Varanasi 221005}
\author{M.B.~Shen}
\affiliation{YaLong River Hydropower Development Company, Chengdu 610051}
\author{L.~Singh}
\altaffiliation{Participating as a member of TEXONO Collaboration}
\affiliation{Institute of Physics, Academia Sinica, Taipei 11529}
\affiliation{Department of Physics, Banaras Hindu University, Varanasi 221005}
\author{M.K. Singh}
\altaffiliation{Participating as a member of TEXONO Collaboration}
\affiliation{Institute of Physics, Academia Sinica, Taipei 11529}
\affiliation{Department of Physics, Banaras Hindu University, Varanasi 221005}
\author{A.K. Soma}
\altaffiliation{Participating as a member of TEXONO Collaboration}
\affiliation{Institute of Physics, Academia Sinica, Taipei 11529}
\affiliation{Department of Physics, Banaras Hindu University, Varanasi 221005}
\author{J. Su}
\affiliation{Key Laboratory of Particle and Radiation Imaging (Ministry of Education) and Department of Engineering Physics, Tsinghua University, Beijing 100084}
\author{C.J. Tang}
\affiliation{College of Physical Science and Technology, Sichuan University, Chengdu 610064}
\author{J.M.~Wang}
\affiliation{YaLong River Hydropower Development Company, Chengdu 610051}
\author{L. Wang}
\affiliation{Key Laboratory of Particle and Radiation Imaging (Ministry of Education) and Department of Engineering Physics, Tsinghua University, Beijing 100084}
\affiliation{College of Physical Science and Technology, Sichuan University, Chengdu 610064}
\author{Q.~Wang}
\affiliation{Key Laboratory of Particle and Radiation Imaging (Ministry of Education) and Department of Engineering Physics, Tsinghua University, Beijing 100084}
\author{S.Y.~Wu}
\affiliation{YaLong River Hydropower Development Company, Chengdu 610051}
\author{Y.C.~Wu}
\affiliation{Key Laboratory of Particle and Radiation Imaging (Ministry of Education) and Department of Engineering Physics, Tsinghua University, Beijing 100084}
\author{Y.C. Wu}
\affiliation{NUCTECH Company, Beijing 100084}
\author{Z.Z. Xianyu}
\affiliation{Key Laboratory of Particle and Radiation Imaging (Ministry of Education) and Department of Engineering Physics, Tsinghua University, Beijing 100084}
\author{R.Q. Xiao}
\affiliation{Key Laboratory of Particle and Radiation Imaging (Ministry of Education) and Department of Engineering Physics, Tsinghua University, Beijing 100084}
\author{H.Y. Xing}
\affiliation{College of Physical Science and Technology, Sichuan University, Chengdu 610064}
\author{F.Z. Xu}
\affiliation{Key Laboratory of Particle and Radiation Imaging (Ministry of Education) and Department of Engineering Physics, Tsinghua University, Beijing 100084}
\author{Y. Xu}
\affiliation{School of Physics, Nankai University, Tianjin 300071}
\author{X.J. Xu}
\affiliation{Key Laboratory of Particle and Radiation Imaging (Ministry of Education) and Department of Engineering Physics, Tsinghua University, Beijing 100084}
\author{T.~Xue}
\affiliation{Key Laboratory of Particle and Radiation Imaging (Ministry of Education) and Department of Engineering Physics, Tsinghua University, Beijing 100084}
\author{C.W.~Yang}
\affiliation{College of Physical Science and Technology, Sichuan University, Chengdu 610064}
\author{L.T. Yang}
\affiliation{Key Laboratory of Particle and Radiation Imaging (Ministry of Education) and Department of Engineering Physics, Tsinghua University, Beijing 100084}
\author{S.W. Yang}
\altaffiliation{Participating as a member of TEXONO Collaboration}
\affiliation{Institute of Physics, Academia Sinica, Taipei 11529}
\author{N. Yi}
\affiliation{Key Laboratory of Particle and Radiation Imaging (Ministry of Education) and Department of Engineering Physics, Tsinghua University, Beijing 100084}
\author{C.X. Yu}
\affiliation{School of Physics, Nankai University, Tianjin 300071}
\author{H. Yu}
\affiliation{Key Laboratory of Particle and Radiation Imaging (Ministry of Education) and Department of Engineering Physics, Tsinghua University, Beijing 100084}
\author{X.Z. Yu}
\affiliation{College of Physical Science and Technology, Sichuan University, Chengdu 610064}
\author{X.H. Zeng}
\affiliation{YaLong River Hydropower Development Company, Chengdu 610051}
\author{Z. Zeng}
\affiliation{Key Laboratory of Particle and Radiation Imaging (Ministry of Education) and Department of Engineering Physics, Tsinghua University, Beijing 100084}
\author{L.~Zhang}
\affiliation{NUCTECH Company, Beijing 100084}
\author{Y.H. Zhang}
\affiliation{YaLong River Hydropower Development Company, Chengdu 610051}
\author{M.G. Zhao}
\affiliation{School of Physics, Nankai University, Tianjin 300071}
\author{W. Zhao}
\affiliation{Key Laboratory of Particle and Radiation Imaging (Ministry of Education) and Department of Engineering Physics, Tsinghua University, Beijing 100084}
\author{Z.Y. Zhou}
\affiliation{Department of Nuclear Physics, China Institute of Atomic Energy, Beijing 102413}
\author{J.J. Zhu}
\affiliation{College of Physical Science and Technology, Sichuan University, Chengdu 610064}
\author{W.B. Zhu}
\affiliation{NUCTECH Company, Beijing 100084}
\author{X.Z. Zhu}
\affiliation{Key Laboratory of Particle and Radiation Imaging (Ministry of Education) and Department of Engineering Physics, Tsinghua University, Beijing 100084}
\author{Z.H. Zhu}
\affiliation{YaLong River Hydropower Development Company, Chengdu 610051}

\collaboration{CDEX Collaboration}
\noaffiliation



\date{\today}

\begin{abstract}
  We report the results of searches for solar axions and galactic dark matter axions or axion-like particles with CDEX-1 experiment at the China Jinping Underground Laboratory, using 335.6 kg-days of data from a p-type point-contact germanium detector. The data are compatible with the background model and no excess signals are observed. 
Limits of solar axions on the model independent coupling $g_{Ae}<2.5\times10^{-11}$ from Compton, bremsstrahlung, atomic-recombination and deexcitation channel and $g^{\text{eff}}_{AN}\times g_{Ae}<6.1\times10^{-17}$ from $^{57}$Fe M1 transition at 90\% confidence level are derived.
Within the framework of the DFSZ and KSVZ models, our results exclude the axion mass heavier than 0.9 eV/c$^{2}$ and 173 eV/c$^{2}$, respectively. The derived constraints for dark matter axions below 1 keV improves over the previous results.
\end{abstract}


\maketitle

\section{Introduction}

The China Dark Matter Experiment (CDEX) pursues direct searches of low mass weakly interacting massive particle (WIMP) and studies of double-beta decay in $^{76}$Ge \cite{2014_Olive,2015_Schumann, 2013_Cushman, 2015_Barabash, *2015_Barabash_2, 2012_Bilenky, *2012_Elliott} toward the goal of a ton-scale germanium detector array \cite{CDEX_introduction} at the China Jinping Underground Laboratory (CJPL) \cite{2010_Kang} . CJPL is located in the Jinping traffic tunnel, Sichuan province, China, with a vertical rock overburden of more than 2400 m, providing a measured muon flux of 61.7 y$^{-1}$ m$^{-2}$ \cite{WuYC2013}. 
A pilot measurement CDEX-0 with a germanium detector array with 20~g target mass, achieving the threshold of 177 eVee (``ee" represents electron equivalent energy), was reported \cite{2014_Liu}. 
CDEX-1 experiment adopted one single module of the p-type point-contact germanium ($p$PCGe) detector with fiducial mass of 915 g \cite{2016_Jiang}. The phase-I of CDEX-1 measurement in the absence of anti-Compton detector and prior to surface suppression based on 14.6 kg-days of data was published with a threshold of 400 eVee \cite{CDEX_1kg_2013}. The phase-II measurements, featuring with an anti-Compton detector and bulk surface discrimination, based on the 53.9 kg-days \cite{2014_Yue} and 335.6 kg-days of data \cite{2016_Zhao} were reported. Both results strongly disfavors the allowed region implied by residual excess events from CoGeNT with an identical detector target.

Quantum chromodynamics (QCD), universally believed to be the best theory describing strong interactions, contains the $\Theta$ term which could explicitly give a rise to a measurable CP-violation such as a large neutron electric dipole moment. The experimental bound is about ten orders of magnitude more stringent \cite{2006_Baker} resulting in an unnaturally small upper limit ($<10^{-10}$) to the $\Theta$ parameter. In order to solve this ``strong CP problem", Peccei and Quinn (PQ) postulated a new spontaneously broken symmetry that naturally and dynamically cancels CP violation in the strong interactions \cite{1977_PQ_PRL, *1977_PQ}. Weinberg \cite{1978_Weinberg} and Wilczek \cite{1978_Wilczek} later proposed that this new symmetry introduces a new pseudoscalar particle similar to neutral pions called axion. 
This original axion with a symmetry-breaking scale of the order of the electroweak scale has been excluded by experiments (see \cite{1987_Kim, *2010_Kim, 2012_Ringwald} and refs therein) whereas ``invisible" axion models such as non-hadronic model DFSZ (Dine-Fischler-Srednicki-Zhitnitskii) \cite{1981_Dine, 1980_Zhitniskiy} and hadronic model KSVZ (Kim-Shifman-Vainstein-Zakharov) \cite{1979_Kim, 1980_Shifman} arising from a higher symmetry-breaking energy scale are still allowed. In addition, axion-like-particles (ALPs) with the similar properties as the QCD axions also have the couplings to electrons ($g_{Ae}$), photons ($g_{A\gamma}$) and nucleons ($g_{AN}$), though do not necessarily solve the strong CP problem. 



Several dark matter (DM) experiments aiming at direct detection of WIMPs have reported the axion searches results \cite{2009_Ahmed, 2001_Bernabei, 2011_Aalseth, 2013_CUORE, 2013_EDELWEISS, 2013_XMASS, 2014_XENON100}. These experiments mainly incorporate two detection mechanisms. The first is that axions from our sun have the couplings to photons ($g_{A\gamma}$) in detectors through the Primakoff effect, $a+Q\to Q+\gamma$ (Q stands for charged particles). These measurements utilize the Bragg diffraction effect in the crystal detectors \cite{2013_EDELWEISS,  2009_Ahmed, 2001_Bernabei} in which the intense electric field would enhance the interaction cross-section. The constraints on $g_{A\gamma}$ from these experiments are typically much less sensitive than the helioscope experiment \cite{2011_CAST} and microwave cavity experiment \cite{2014_ADMX}. 
The second is that solar axions and dark matter ALPs have the couplings to electron ($g_{Ae}$) in detectors through the axioelectric effect:
\begin{eqnarray}
\label{eq_axioelectric}
a+e+Z\to e+Z
\end{eqnarray}
which is  similar to the photoelectric effect with the absorption of an axion instead of a photon \cite{2009_Ahmed, 2011_Aalseth, 2013_CUORE, 2013_EDELWEISS, 2013_XMASS, 2014_XENON100}. 

We report the axion searches results from the CDEX-1 experiment based on an exposure of 335.6 kg-days of data which is the same data set as the ref.~\cite{2016_Zhao}. We focus on the $g_{Ae}$ couplings from the solar axions and the galactic dark matter ALPs through axioelectric effect of Eq.~\ref{eq_axioelectric}. 
Studies on the $g_{A\gamma}$ coupling are not pursued since the Bragg diffraction methods are less sensitive and with larger systematic uncertainties.

\section{Axion searches with CDEX-1}

\subsection{CDEX-1 setup and overview}
CDEX-1 experiment adopted one single module of the p-type point-contact germanium ($p$PCGe) detector at 994 g of mass \cite{CDEX_1kg_2013, 2014_Yue, 2016_Zhao}, featuring with a relative low threshold down to 475 eVee. 
A cylindrical NaI(Tl) crystal with well shaped cavity enclosing the cryostat of the $p$PCGe, whose threshold was about 5 keVee, was served as the anti-Compton detector. Events in coincidence with the AC detector were discarded to get rid off $\gamma$-ray induced background.

The $p^{+}$ point-contact electrode after a pulsed-reset feedback preamplifier generated three identical energy-related signals. These three outputs were fed into the shaping amplifiers at 6 $\mu$s (S$_{p6}$), 12 $\mu$s (S$_{p12}$) shaping time, and a timing amplifier (T$_{p}$) respectively. The outputs from S$_{p6,12}$ provided the energy measurement and system trigger of the DAQ. Their dynamic ranges were limited to 12 keVee to achieve the maximal signal-to-noise ratio and maximal information for low-energy events. The output from T$_{p}$ recording the raw fast pulse shape was employed to discriminate the bulk/surface events. Its energy dynamic range could be extended to 20 keVee, while it was slightly different from S$_{p6,12}$ below 2 keVee and had higher energy threshold. So in our following analysis, the spectrum below 12 keVee was from S$_{p6}$ and above 12 keVee was from T$_{p}$. Both calibrations with good linearity of less than 0.8\% deviation were derived from the internal cosmogenic x-ray peaks and random trigger events \cite{2016_Zhao}.

\subsection{Axion sources}
\subsubsection{Solar Axions}
The Sun can be an abundant source of axions, which are generated by four production mechanisms that depend on $g_{Ae}$ \cite{2013_CBRD}:
\begin{enumerate}[(i)]
\item Compton-like scattering (C): $\gamma+e\to e+a$
\item Axion-bremsstrahlung (B): $e+Q\to e+Q+a$
\item Atomic-recombination (R): $e+I\to I^{-}+a$
\item Atomic-deexcitation (D): $I^{*}\to I+a$
\end{enumerate}
where $Q$ is any charged particle, $e$ is electron, $I$ is ion and $I^{*}$ is its excited state.

The fluxes of CBRD processes are estimated by \cite{2013_CBRD,2013_Redondo} and as to the CB solar axions, the fluxes are
\begin{eqnarray}
\label{eq3}
&&\frac{d\Phi_{\text{CB}}}{dE_{A}}=\frac{d\Phi_{\text{C}}}{dE_{A}}+\frac{d\Phi_{\text{B}}}{dE_{A}}\nonumber\\
&&=g^{2}_{Ae}\times1.33\times10^{33}E_{A}^{2.987}e^{-0.776E_{A}}\nonumber\\
&&+g^{2}_{Ae}\times2.63\times10^{35}E_{A}e^{-0.77E_{A}}\frac{1}{1+0.667E_{A}^{1.278}}
\end{eqnarray}
where the unit is $\text{cm}^{-2}\text{s}^{-1}\text{keV}^{-1}$ and axion energy $E_{A}$ is in keV. For RD solar axions, the flux also depend on $g_{Ae}^{2}$ and the tabulated spectrum in ref.~\cite{2013_Redondo} is adopted.

\begin{figure}[htb]
  \includegraphics[width=1.0\linewidth]{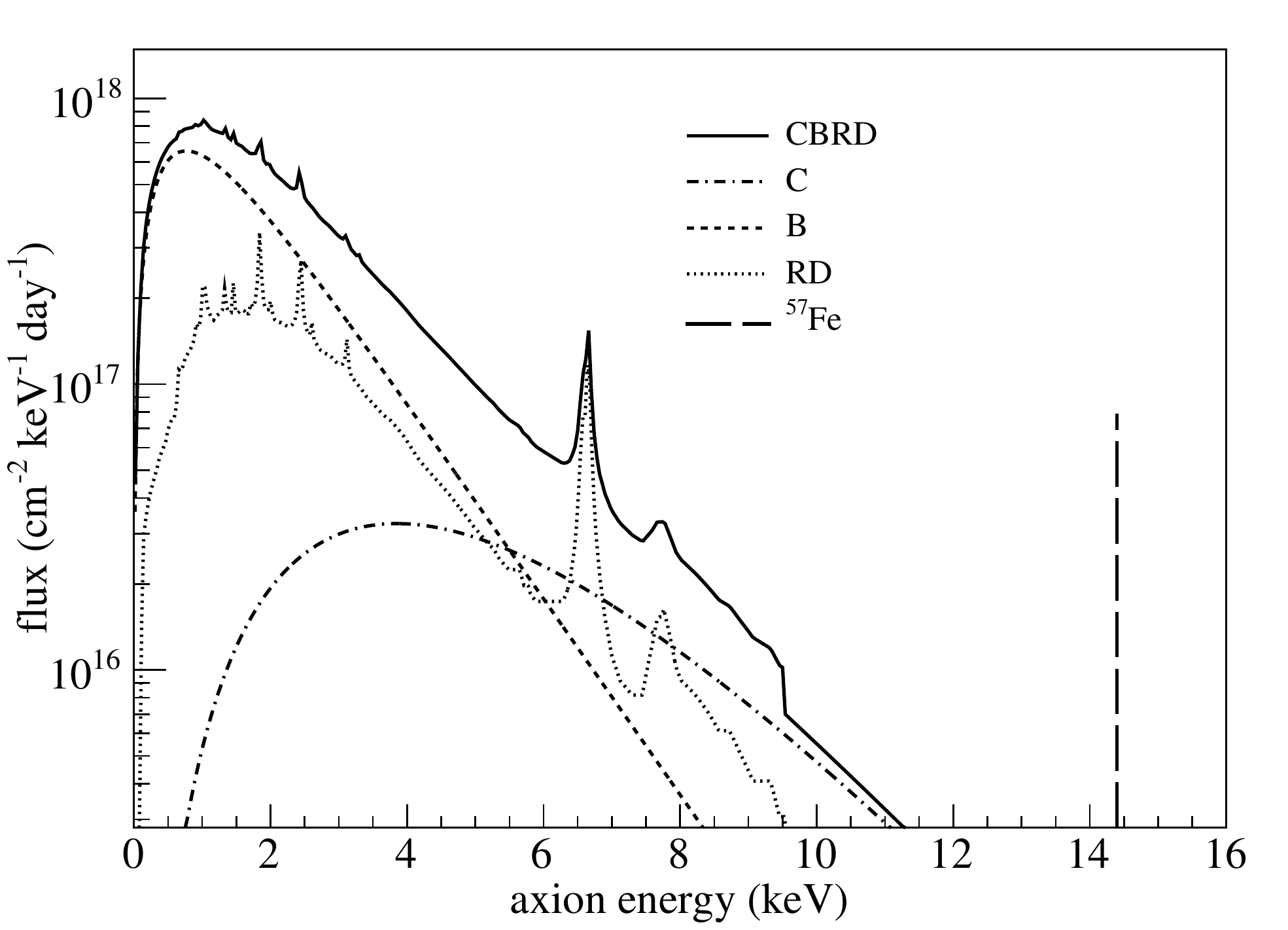}
  \caption{\label{fig:Fig1_flux} The predicted fluxes of solar axions on Earth from different processes. The thick solid line: the combination of Compton (C), bremsstrahlung (B), atomic-recombination (R) and atomic-deexcitation (D) mechanisms. The thick dash line: $^{57}$Fe M1 transition with intrinsic Doppler broadening at the mass of zero. The corresponding axion coupling constants are $g_{Ae}=10^{-11}$ and $g^{\text{eff}}_{AN}=10^{-7}$.}
\end{figure}

The 14.4 keV monochromatic axions emitted in the M1 transition of $^{57}$Fe nuclei in the Sun
\begin{eqnarray}
\label{eq_Fe57}
^{57}\text{Fe}^{*}\to^{57}\text{Fe} + a
\end{eqnarray}
can be an additional important source of solar axions due to the large abundance of $^{57}$Fe among the heavy elements \cite{1991_Fe57, 1995_Fe57}.
Its flux is related to $g_{AN}$ coupling and is given by ~\cite{2009_Andriamonje,2013_EDELWEISS}.

\begin{eqnarray}
\label{eq4}
\Phi_{14.4}=(\frac{k_{A}}{k_{\gamma}})^{3}\times4.56\times10^{23}(g^{\text{eff}}_{\text{AN}})^{2}
\end{eqnarray}
where the unit is $\text{   cm}^{-2}\text{s}^{-1}$, $k_{A}$ and $k_{\gamma}$ are the momenta of the outgoing axion and photon respectively. The effective nuclear coupling $g^{\text{eff}}_{AN}$ is model dependent, $g^{\text{eff}}_{AN}\equiv(-1.19g^{0}_{AN}+g^{3}_{AN})$, where $g^{0}_{AN}$ and $g^{3}_{AN}$ are the model-dependent isoscalar and isovector axion-nucleon coupling constants, respectively ~\cite{1985_Kaplan,1985_Srednicki_}. 
Fig.~\ref{fig:Fig1_flux} shows the evaluated fluxes of solar axions on Earth for the processes we are concerned.

\subsubsection{Galactic Dark matter ALPs}
The non-thermal axions or ALPs, produced by the vacuum realignment mechanism and radiation from cosmic strings, are candidates to solve the dark matter problem in the universe.
The total average flux independent on any axion coupling is given by
\begin{eqnarray}
\label{eq5}
\Phi_{\text{DM}}&&=\rho_{\text{DM}}\cdot v_{A}/m_{A}\nonumber\\
&&=9.0\times10^{15}\times\beta\cdot(\frac{\text{keV}}{m_{A}}) \text{ cm}^{-2}\text{s}^{-1}
\end{eqnarray}
where $\rho_{\text{DM}}$ is the dark matter halo density ($\rho_{\text{DM}}\sim0.3$ GeV/cm$^{3}$ ~\cite{2012_Green}), $m_{A}$ is the axion mass, $v_{A}$ is the mean axion velocity distribution with respect to the Earth, $\beta$ is the ratio of the axion velocity to the speed of light. 


\subsection{Experimental Signatures}
We focus on the detection channel of axioelectric effect as illustrated in Eq.\ref{eq_axioelectric}, where the cross-section $\sigma_{Ae}$ is given by
\begin{eqnarray}
\label{eq2}
\sigma_{Ae}(E_{A})=\sigma_{pe}(E_{A})\frac{g^{2}_{Ae}}{\beta}\frac{3E^{2}_{A}}{16\pi\alpha m^{2}_{e}}(1-\frac{\beta^{\frac{2}{3}}}{3})
\end{eqnarray}
as described in \cite{2013_CUORE, 2010_Derevianko, 2008_Pospelov}, 
where $\sigma_{pe}(E_{A})$ is the photoelectric cross section for Ge, $\alpha$ is the fine structure constant, $m_{e}$ is the electron mass and $\beta$ is the ratio of the axion velocity to the speed of light.

The expected axion event rate at measurable energy $E$ is obtained by the convolution of the flux, the axioelectric cross section and the energy resolution of the detector:
\begin{eqnarray}
\label{eq1}
R_{i}(E&)&=\int{dE_{A}\sigma_{Ae}(E_{A})(\frac{d\Phi_{i}}{dE_{A}})\times\frac{1}{\sqrt{2\pi}\sigma}e^{-\frac{(E-E_{A})^{2}}{2\sigma^{2}}}}
\end{eqnarray}
where $i$ represents the different axion sources of fluxes $\Phi_{i}$. The detector energy resolution ($\sigma$) is 90 eV at 10.37 keV \cite{2016_Zhao}. 

\begin{figure}[htb]
  \includegraphics[width=1.0\linewidth]{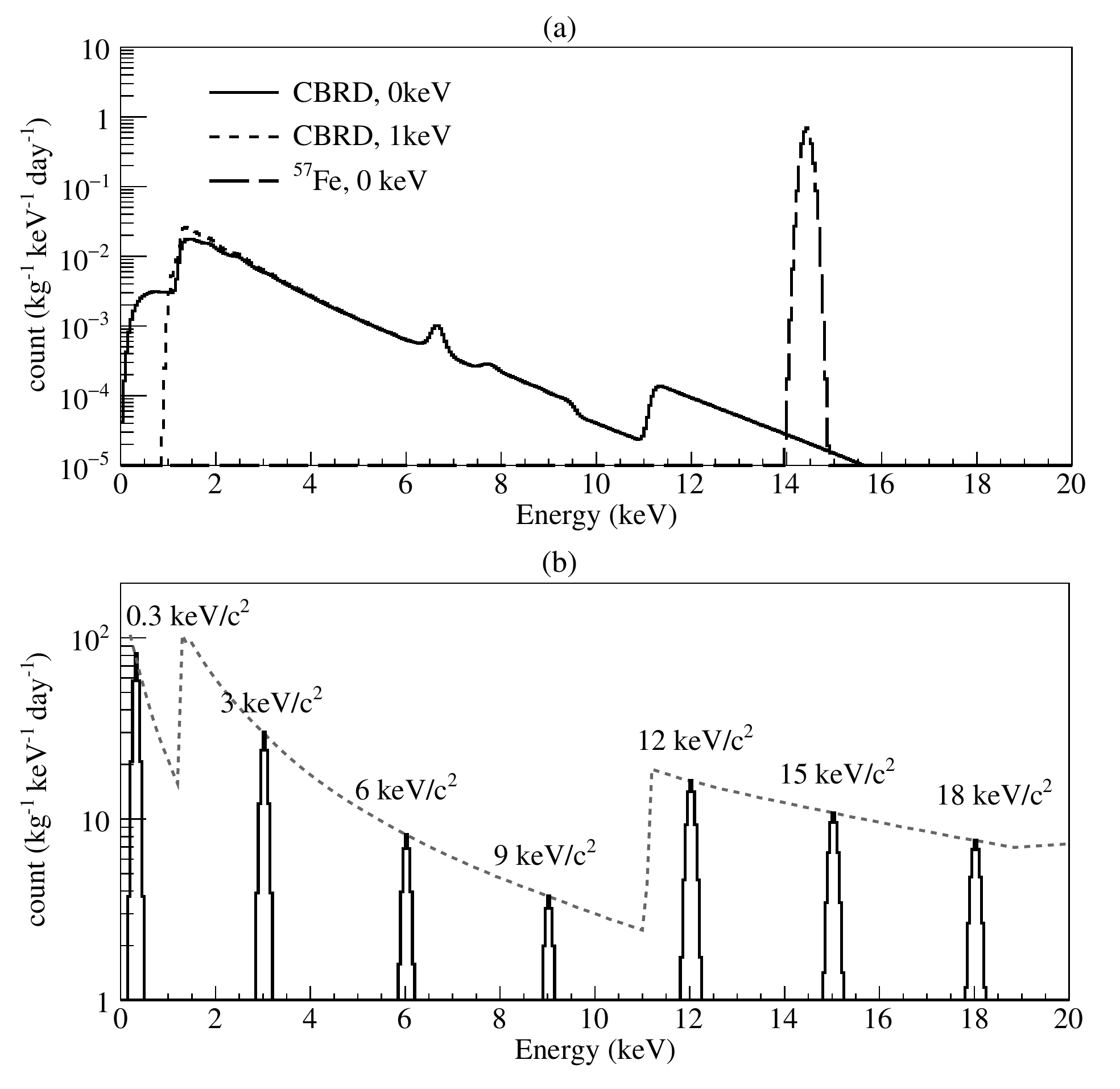}
  \caption{\label{fig:Fig2_expected_rates} The expected axion event rates in CDEX-1 detector with energy resolution. (a) solar axions: solid line, CBRD axion at the mass close to 0 keV; dotted line, CBRD axion at the mass of 1 keV; dashed line, $^{57}$Fe 14.4 keV axion at the mass of 0 keV.
  (b) The dashed line is the maximum event rate of DM ALPs Gaussian distributions versus their masses; The signal signatures as shown in solid lines are Gaussian distributions with width determined by energy resolution and the maximum point described by the dashed line. Here the axion couplings are $g_{Ae}=10^{-11}$ and $g^{\text{eff}}_{AN}\times g_{Ae}=10^{-16}$.}
\end{figure}

In particular, since the ALP DM is cold ($\beta\approx10^{-3}$), Eq.~\ref{eq1} translates to \cite{2008_Pospelov}:
\begin{eqnarray}
\label{eq_DM_rate}
R=1.2\times10^{43}A^{-1}g^{2}_{Ae}m_{A}\sigma_{pe}(m_{A})
\end{eqnarray}
where A is mass number for germanium and the units of $R$ and $\sigma_{pe}$ are kg$^{-1}$day$^{-1}$ and barns/atom, respectively.
We note that the sensitivity dependence on the coupling strength are different for different sources and detector channels. The event rate $R_{i}(E)$ varies as $g_{Ae}^{4}$, $(g_{Ae}\times g^{\text{eff}}_{\text{AN}})^{2}$, $g_{Ae}^{2}$ for $i=$CBRD, $^{57}$Fe, DM, respectively. The difference in coupling dependence of ALP DM rates compared to those of solar axions is a consequence of the DM flux being fixed by cosmological data given a certain $m_{A}$.
The expected $R_{i}(E)$ for various channels in CDEX-1 are depicted in Fig.~\ref{fig:Fig2_expected_rates}.

\section{DATA ANALYSIS}
\subsection{Candidate Event Selection}
The background spectrum is derived by the same selection procedures used in earlier analysis \cite{2014_Yue,2016_Zhao}:

(i) Stability check, which discards the time periods of calibration or laboratory construction;

(ii) Physics versus electronic noise, which differentiates physical events from the electronic noise and spurious signals. 

(iii) Anti-Compton selection, which removes the events in coincidence with the anti-Compton detector. 

In particular, there exists an inactive layer of about 1 mm in thickness at the $n^{+}$ surface electrode. These surface events are rejected by pulse shape analysis using their characteristic slower rise-time \cite{2014_Yue,2016_Zhao}. Procedures have been established to derive their signal-retaining and background-leakage efficiencies \cite{TEXONO_2013,*2014_LHB}.

\begin{figure}[htb]
  \includegraphics[width=1.0\linewidth]{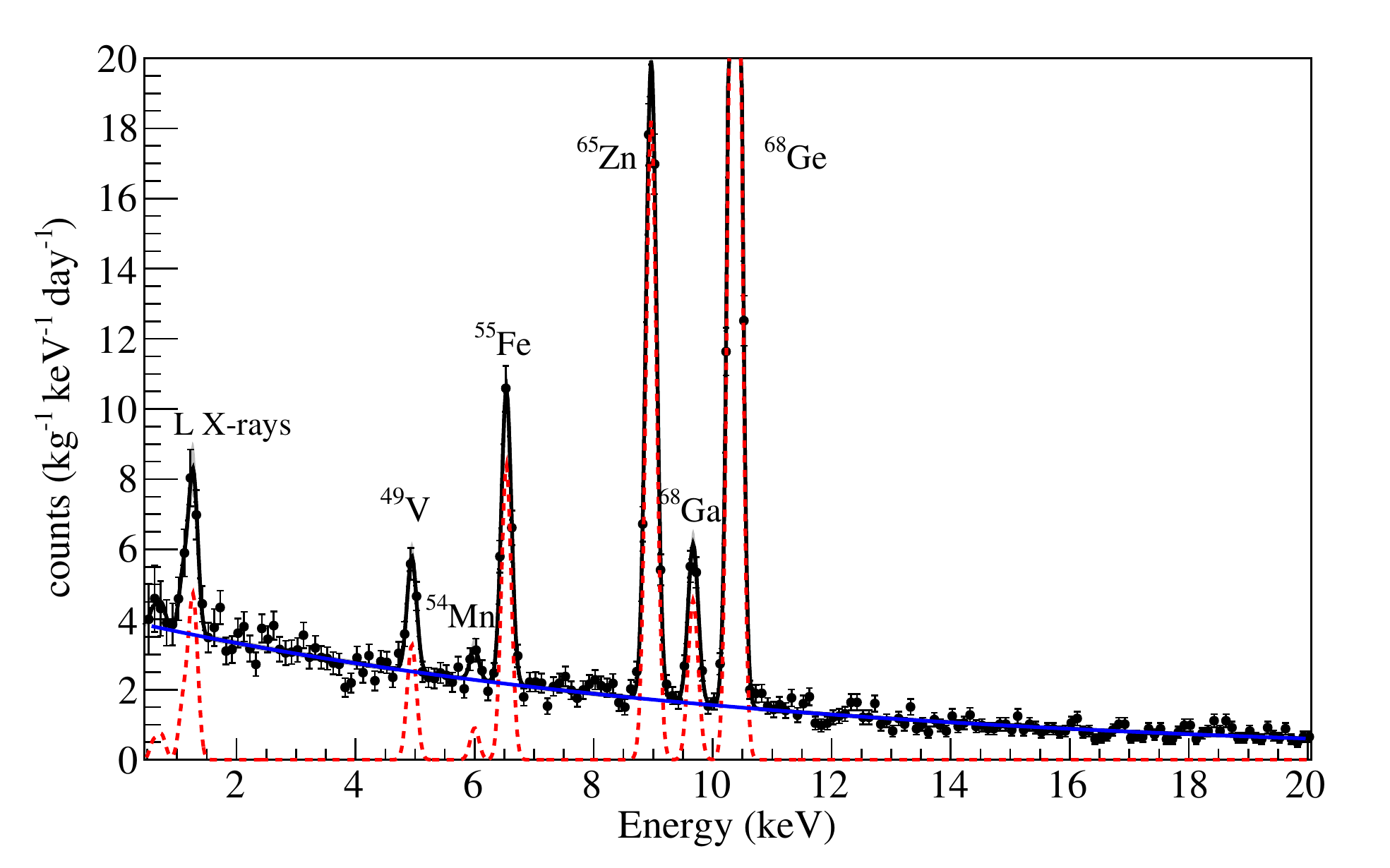}
  \caption{\label{fig:Fig1_Br} (color online) The efficiency corrected bulk spectrum $B_{0}$ (black dots) from 475 eV up to 20 keV, as well as the background assumption of K/L shell x-rays (dotted red line) and continuous background (solid blue line).}
\end{figure}


\subsection{Background description and background model simplification}

In this work, we analyze the same data set as the ref.\cite{2016_Zhao} with an exposure of 335.6 kg-days of data. The bulk spectrum ($B_{0}$) from 475 eV up to 20 keV after the data selection described above and after efficiency correction is shown in Fig.~\ref{fig:Fig1_Br}. The background consists of six distinct K-shell x-ray peaks from the cosmogenic nuclides and their corresponding L-shell x-rays (dotted red line), and a continuous component  with a smooth, slightly increasing profile as the energy decreasing (solid blue line)\cite{2014_Yue,2016_Zhao}. 

The contribution of the ambient radioactivity at CJPL external to the shielding system was greatly suppressed to $\ll10^{-6}$~kg$^{-1}$ keV$^{-1}$ day$^{-1}$ at the energy range below 20 keV \cite{2016_Zhao}. The continuous background below 20 keV was expected to mainly originate from residual $^{238}$U, $^{232}$Th, $^{40}$K in the experimental hardware in the vicinity of the $p$PCGe detector, radon gas penetrating inside shielding, and cosmogenic $^{3}$H inside the crystal. Quantitative studies of their relative contributions are our current research efforts and beyond the scope of this work. 

However all the axion signals have the signatures which are significantly different from the continuous background especially in the local energy range. 
As shown in Fig.~\ref{fig:Fig2_expected_rates}, for $^{57}$Fe solar axions and dark matter ALPs, compared with continuous background, these event signatures are monochromatic and Poisson distributions whose widths are determined by the energy resolution. As to the continuous CBRD solar axion, the event rate have the distinct signature that it has a saw-tooth-like profile between the local energy from 0.9 keV to 1.6 keV if we only consider the mass within 1 keV/c$^{2}$. As discussed in ref.\cite{2013_Redondo}, the most recent and accurate calculation for solar axion flux is valid for light axions, hence we only consider the axion mass $m_{A}<1 \text{ keV/c}^{2}$. 

The accurate quantitative study of the continuous background is not essential for this axion sensitivity experiment.
Therefore we interpret the background in a simplified way: the combination of K/L x-ray peaks and a continuous background. 
A constant background within a local energy range of interest is sufficient for this analysis. The formulation of the analysis algorithms and evaluation of systematic uncertainties are discussed in subsequent sections.

\subsection{Analysis method}
The unbinned maximum likelihood method ~\cite{2009_Ahmed,2014_XENON100} is adopted to derive constraints in axion couplings from the measured spectrum. 
Every measured event is categorized as bulk or surface event, denoted by $B_{m}$ and $S_{m}$, respectively, according to its rise time.
The relationships between the measured spectrum ($B_{m}$, $S_{m}$) and the efficiency corrected spectrum ($B_{0}$, $S_{0}$) can be derived from the following coupled equations, which are illustrated in Ref.~\cite{TEXONO_2013,*2014_LHB,2014_Yue,2016_Zhao}:
\begin{eqnarray}
\label{eq8}
B_{m}=\varepsilon_{BC}\cdot[\varepsilon_{BS}\cdot B_{0}+(1-\lambda_{BS})\cdot S_{0}]\nonumber\\
S_{m}=\varepsilon_{BC}\cdot[\lambda_{BS}\cdot S_{0}+(1-\varepsilon_{BS})\cdot B_{0}]
\end{eqnarray}
$\varepsilon_{BC}$ refers to the combined efficiencies of physics versus electronic noise selection and Anti-Compton selection mentioned in Sec.III.A. 
The efficiencies of $\varepsilon_{BS}$ and $\lambda_{BS}$, representing the bulk event retaining and surface background rejection, can translate ($B_{m}$, $S_{m}$) to ($B_{0}$, $S_{0}$). 

\begin{figure}[htb]
  \includegraphics[width=1.0\linewidth]{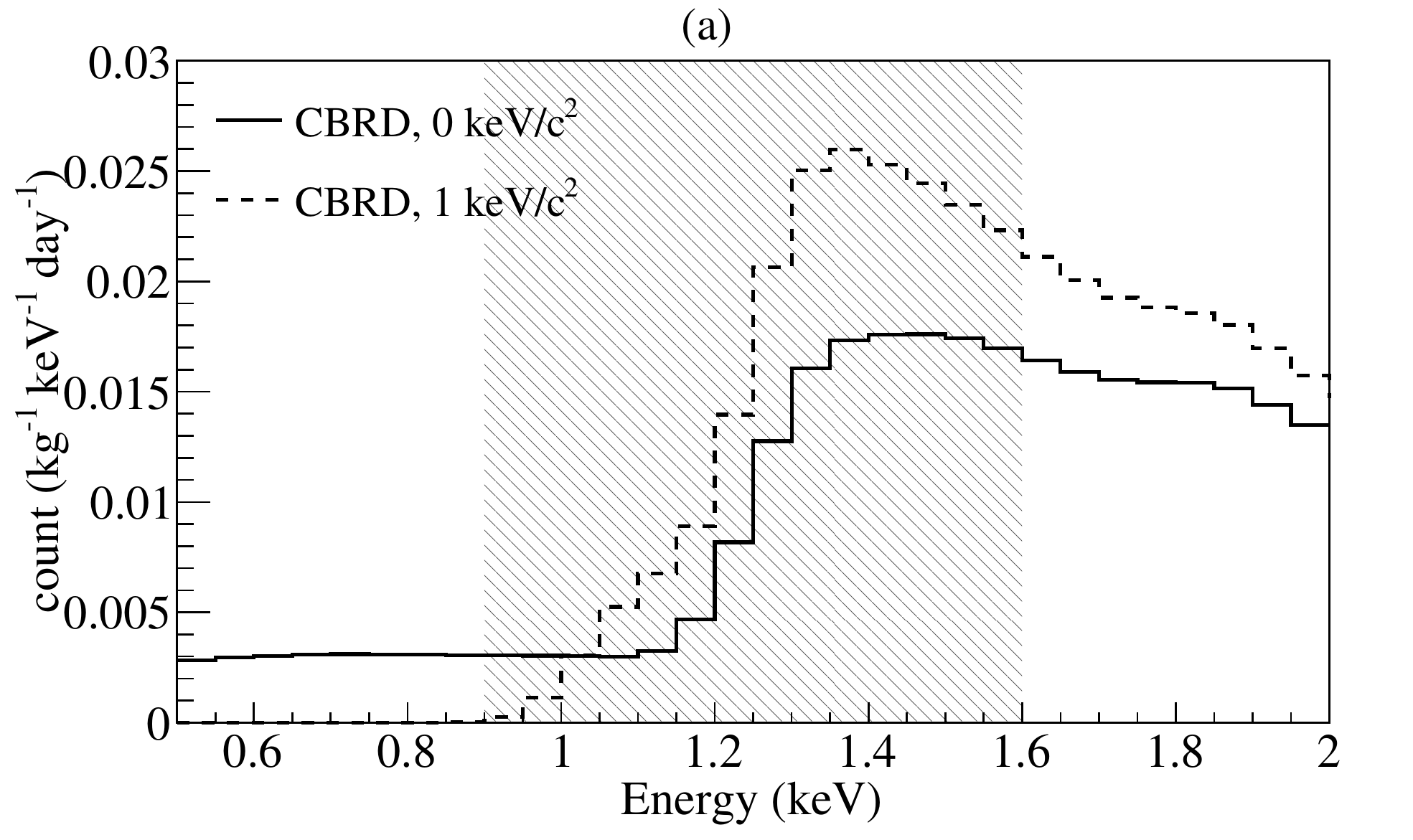}
    \includegraphics[width=1.0\linewidth]{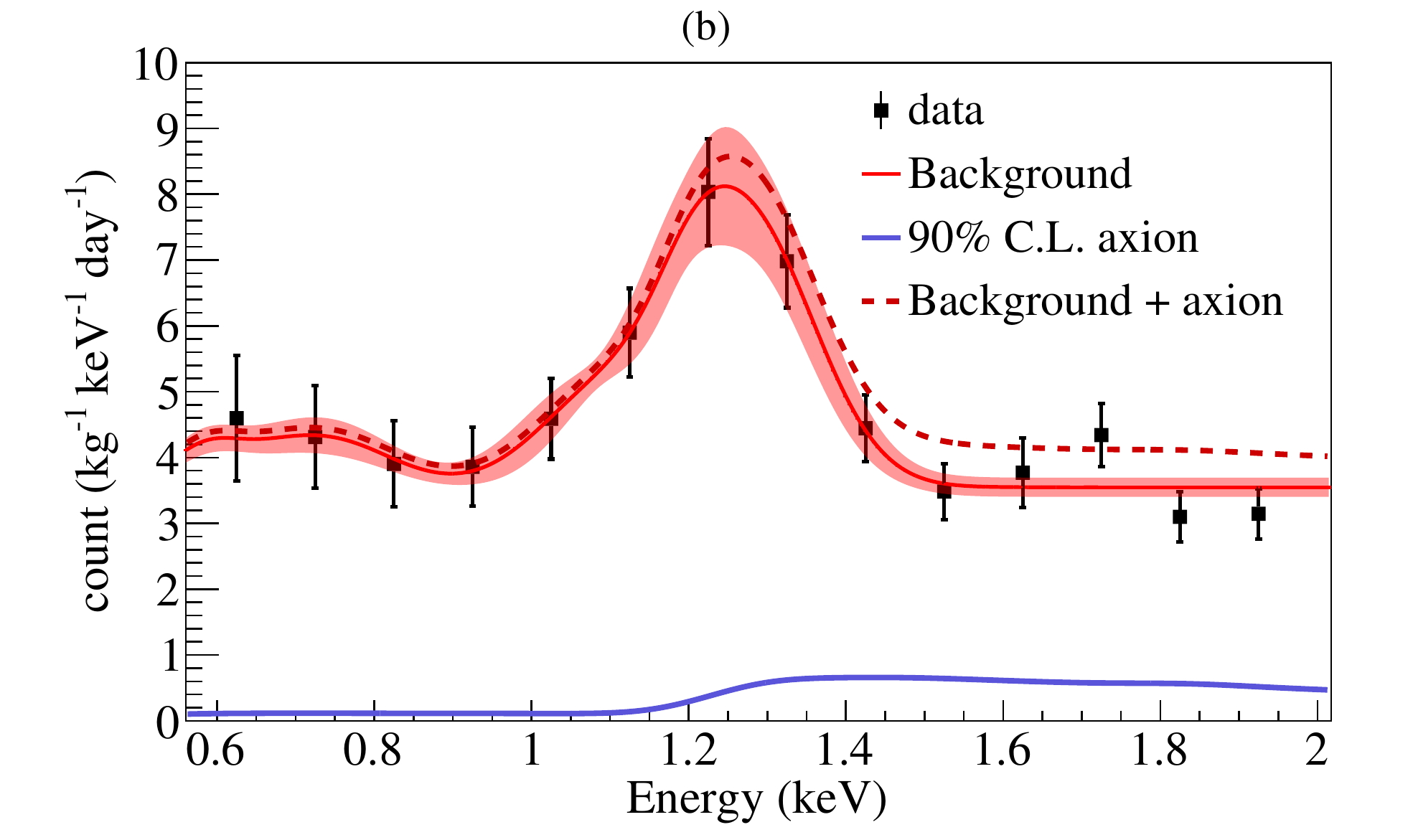}
  \caption{\label{fig:C_B_RD} (a) The expected axion rates in CDEX-1 detector at the mass of 0 keV/c$^{2}$ and 1 keV/c$^{2}$; (b) The 90\% C.L. CBRD axion result at mass of zero (blue line) and the bulk data (black data points) in 0.475 -- 2 keV energy range, as well as the background assumption (red line) and background + 90\% C.L. axion signal (dashed red line).}
\end{figure}

The best-fit solution to ($B_{0}$ and $S_{0}$ is evaluated by maximizing the likelihood function \cite{1998_statistical}:

\begin{eqnarray}
\label{eq_ML}
\mathcal{L}=\prod_{i=1}^{N_{B_{m}}}P_{B_{m_{i}}}\cdot\prod_{j=1}^{N_{S_{m}}}P_{S_{m_{j}}}
\end{eqnarray}
where $N_{B_{m}}$ and $N_{S_{m}}$ are the numbers of bulk measurement events and surface measurement events respectively. $P_{B_{m}}$ and $P_{S_{m}}$ are the p.d.f.s (probability density function) of bulk measurement and surface measurement respectively, which are described by

\begin{eqnarray}
\label{eq7}
P_{B_{m}}&&=\varepsilon_{BC}\cdot[\varepsilon_{BS}\cdot(\alpha_{local}\cdot P_{local}+\alpha_{K,L}\cdot P_{K,L}\nonumber\\
&&+\alpha_{A}\cdot P_{A})+(1-\lambda_{BS})\cdot\alpha_{S_{0}}P_{S_{0}}]\\
P_{S_{m}}&&=\varepsilon_{BC}[(1-\varepsilon_{BS})\cdot(\alpha_{local}^{'}\cdot P_{local}+\alpha_{K,L}^{'}\cdot P_{K,L}\nonumber\\
&&+\alpha_{A}^{'}\cdot P_{A})+\lambda_{BS}\cdot\alpha_{S_{0}}^{'}P_{S_{0}}]
\end{eqnarray}
According to previous discussion of simplified background model, the first background component: $P_{local}$ represents the normalized p.d.f. of local continuous background using a zeroth polynomial function. The fitting range is constrained to local by different kinds of  axion sources. As to CBRD solar axion, the fitting range is limited to 0.9 keV to 1.6 keV as shown in Fig.\ref{fig:C_B_RD} (a); for $^{57}$Fe, that is limited to 13.0 keV to 16.0 keV as depicted in Fig.\ref{fig:Fig8_Fe57}; and to ALP DM axion, the range is constrained to $\pm$8$\sigma$ range.
The other component: $P_{K,L}(E)$ is the normalized p.d.f of the K/L shell x-rays peaks. $P_{A}$ is the normalized p.d.f. describing the axion events as shown in Fig.~\ref{fig:Fig2_expected_rates}. $P_{S_{0}}$ represents the normalized p.d.f. of the efficiency corrected surface spectrum $S_{0}$, derived  from fitting $S_{0}$ by a smooth curve. The systematic uncertainties of the p.d.f. selection of $P_{S_{0}}$ is negligible by comparing bin-by-bin p.d.f from $S_{0}$ spectrum.

The relative contributions of each components in the bulk measurements are represented by $\alpha_{local}$, $\alpha_{K,L}$, $\alpha_{A}$ and $\alpha_{S_{0}}$, while $\alpha_{local}^{'}$, $\alpha_{K,L}^{'}$, $\alpha_{A}^{'}$ and $\alpha_{S_{0}}^{'}$ are the their individual relative contributions to the surface measurement data.
The measured bulk and surface components are related by
\begin{eqnarray}
\label{eq_*}
\frac{\alpha_{*}}{\alpha^{'}_{*}}=\frac{\int\varepsilon_{BC}\varepsilon_{BS}P_{*}dE}{\int\varepsilon_{BC}(1-\varepsilon_{BS})P_{*}dE}
\end{eqnarray}
where * represents $local, K/L$ and $A$. In addition, the efficiency-corrected $S_{0}$ component is given by
\begin{eqnarray}
\label{eq_n_n*}
\frac{\alpha_{S_{0}}}{\alpha^{'}_{S_{0}}}=\frac{\int\varepsilon_{BC}(1-\lambda_{BS})P_{S_{0}}dE}{\int\varepsilon_{BC}\lambda_{BS}P_{S_{0}}dE}
\end{eqnarray}
The goodness-of-fit of this maximum likelihood analysis is tested with binned the data, where $\chi^{2}/n_{d}=1.2$ ($n_{d}$ represents the degrees of freedom) at the energy of 300~eV of DM ALPs.

\subsection{Systematic Uncertainties}

The effects of systematic uncertainties have been evaluated for all of analyses from the three different axion sources.
Systematic uncertainties may originate from bulk surface events selection, signal selection, fiducial mass as well as the background assumption.

According to the evaluation in the previous work \cite{2016_Zhao}, the contribution of the uncertainty of bulk surface event selection in the low energy range is dominated. 
This component has been taken into account in the likelihood function of Eq.~\ref{eq_ML}, and introduced via the uncertainties of $\varepsilon_{BS}$ and $\lambda_{BS}$ in Eq.~\ref{eq8}. 
This contributes about 55\%, 15\% and well below 1\% systematic uncertainties to the constraints on galactic dark matter axion below 1 keV/c$^{2}$, CBRD solar axion and $^{57}$Fe solar axion, respectively.

The uncertainties of the background assumption $P_{local}$ have been evaluated by the different background assumptions between the flat background, polynomial background and exponential background.
The variation of the background assumptions gives the uncertainties of about 7\%, 5\% and 8\% for the galactic dark matter axion below 1 keV/c$^{2}$, CBRD solar axion and $^{57}$Fe solar axion, respectively.

Compared to the uncertainties arisen from bulk surface events selection and background assumption, the contributions of signal selection, fiducial mass and energy resolution uncertainties of the detector to the systematic uncertainties are negligible.

\begin{figure}[htb]
  \includegraphics[width=1.0\linewidth]{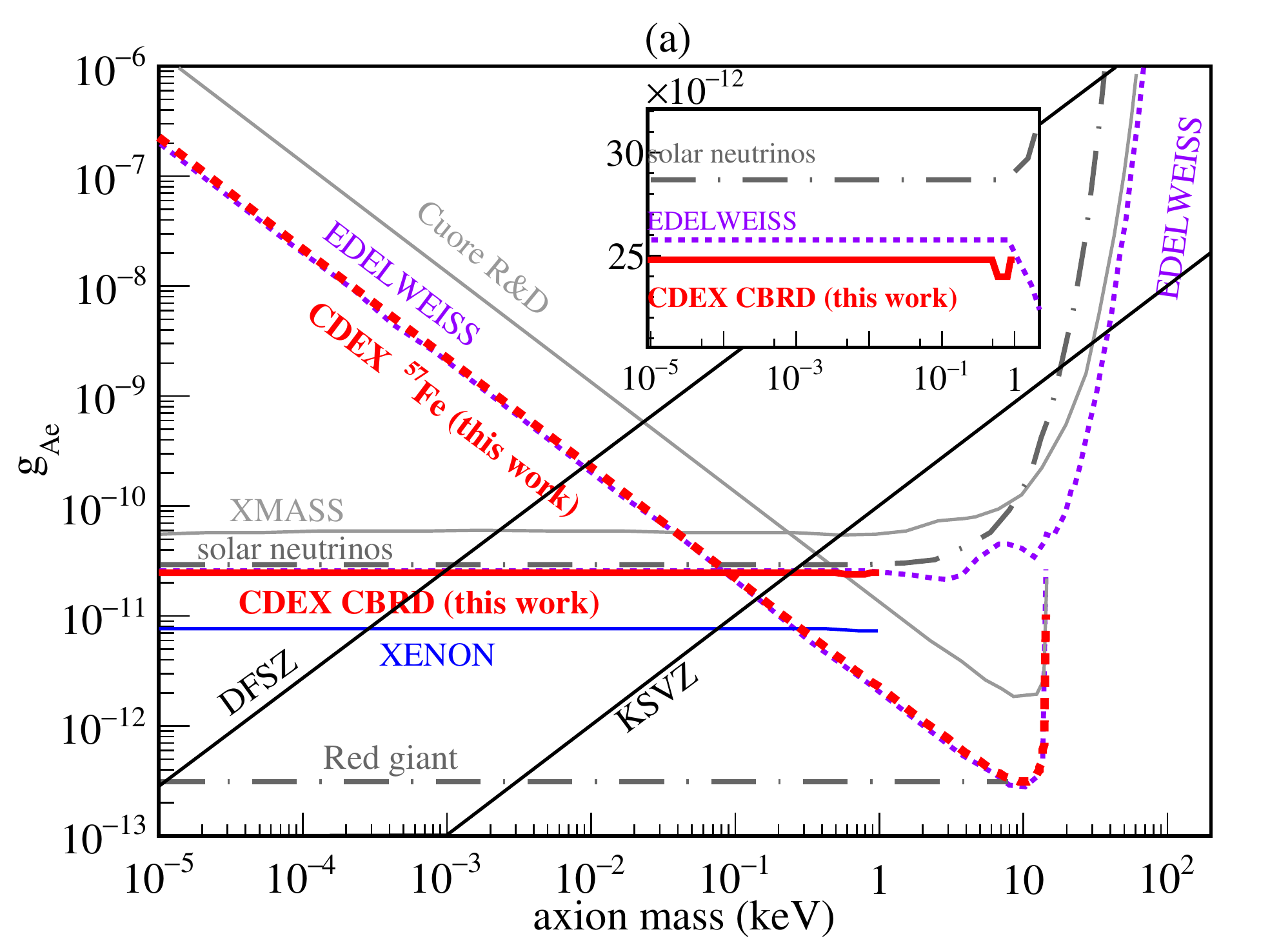}
      \includegraphics[width=1.0\linewidth]{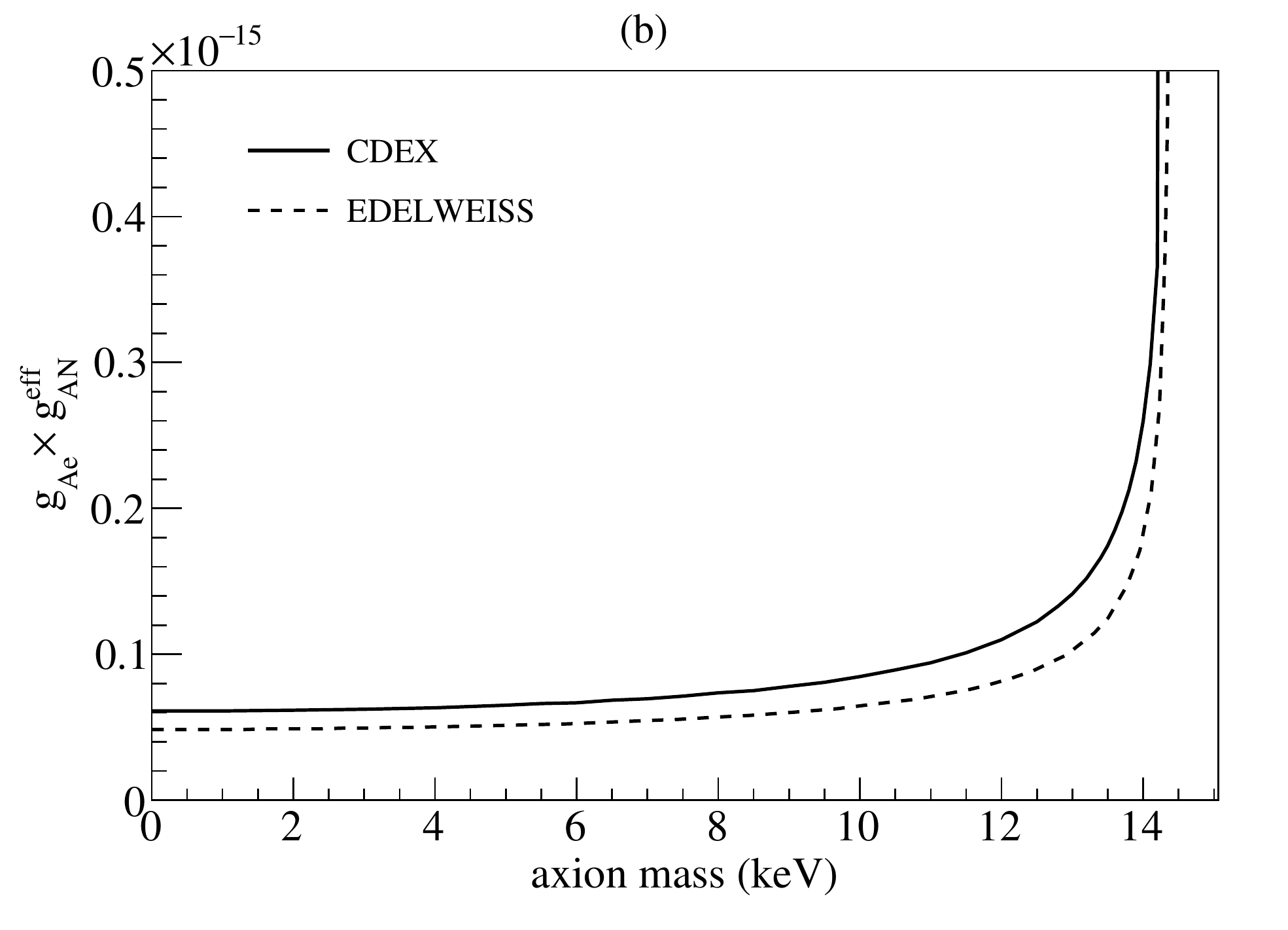}
  \caption{\label{fig:Fig7_Fe57_CB_RD_limit} (a) The CDEX-1 90\% C.L. on CBRD solar axion (solid red line) and $^{57}$Fe 14.4 keV solar axion (dashed red line) in DFSZ model with $S=0.5$ and cos$\beta^{2}_{DFSZ}=1$, together with the bounds from astrophysical bounds \cite{2009_Gondolo,2013_Viaux,2008_Raffelt}, others including CBRD axion and $^{57}$Fe axion \cite{2013_EDELWEISS,2014_XENON100,2013_XMASS,2013_CUORE}. The benchmark DFSZ (cos$\beta^{2}_{DFSZ}=1$) and KSVZ (E/N=0) models are displayed as two solid black lines; (b) The CDEX-1 90\% C.L. on the model independent coupling of  $g^{\text{eff}}_{AN}\times g_{Ae}$ of $^{57}$Fe 14.4 keV solar axion, compared with the EDELWEISS results \cite{2013_EDELWEISS}. }
\end{figure}

\begin{figure}[htb]
  \includegraphics[width=1.0\linewidth]{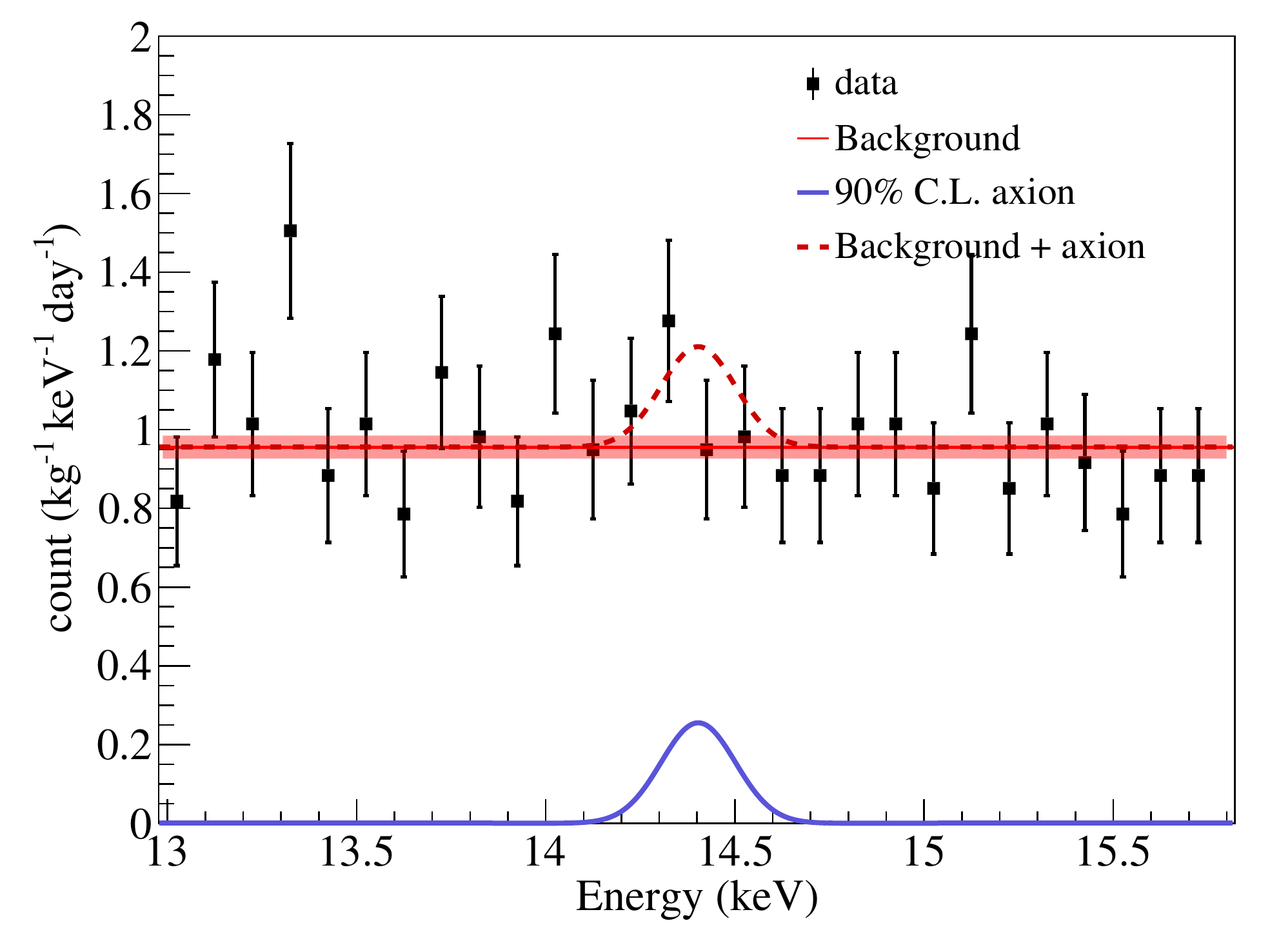}
  \caption{\label{fig:Fig8_Fe57}  The bulk real data (black data points) and the 90\% C.L. $^{57}$Fe axion result (blue line)  in 13 -- 16 keV energy range, as well as the background assumption (red line) background + 90\% C.L. axion signal (dashed red line). }
\end{figure}

\section{RESULTS}
\subsection{Solar axions}
\subsubsection{CBRD}
For the CBRD solar axion, the saw-tooth signature is within the energy range as the shadow displayed in Fig.~\ref{fig:C_B_RD} (a), since we only consider the mass below 1 keV/c$^{2}$.
The local fitting range is limited to the shadow region 0.9~keV $\sim$ 1.6 keV and the fitting results of 90\% C.L. at the mass of 0 keV is shown in Fig.~\ref{fig:C_B_RD} (b), as well as the $B_{0}$ spectrum and the background model described.
The data are compatible with the background model and no excess signals are observed. 
 The solid red line in Fig.~\ref{fig:Fig7_Fe57_CB_RD_limit} (a) shows our limit on $g_{Ae}$ at 90\% C.L. which is restricted to the mass below 1 keV/c$^{2}$, together with the bounds from astrophysical bounds \cite{2009_Gondolo,2013_Viaux,2008_Raffelt}, other representative experiments including CBRD axion and $^{57}$Fe axion \cite{2013_EDELWEISS,2014_XENON100,2013_XMASS,2013_CUORE}. 
 
 As illustrated in the inset of Fig.~\ref{fig:Fig7_Fe57_CB_RD_limit} (a), the improved energy threshold of CDEX-1 gives rise to a 90\% C.L. limit of $2.5\times10^{-11}$ for $g_{Ae}$, which is comparable to that of EDELWEISS experiment \cite{2013_EDELWEISS} which also adopts germanium detectors.

As to a specific axion model, DFSZ or KSVZ \cite{1981_Dine, 1980_Zhitniskiy,1979_Kim, 1980_Shifman}, the $g_{Ae}$ limit can be translated into the limit of axion mass $m_{A}$. In the DFSZ model, on the assumption of model-dependent parameter cos$\beta_{DFSZ}=1$, where $\beta_{DFSZ}$ is an arbitrary angle, CDEX-1 excludes axion masses above 0.9 eV/c$^{2}$. In the KSVZ model, on the assumption of model-dependent parameters $E/N=0$, where $E/N$ is the ratio of the electromagnetic to color anomalies of the Peccei-Quinn symmetry \cite{1985_Srednicki_}, our result excludes axion masses above 265 eV/c$^{2}$.

\subsubsection{$^{57}$Fe}
For $^{57}$Fe M1 transition 14.4 keV axion, Fig.~\ref{fig:Fig8_Fe57} displays the $B_{0}$ spectrum at the energy range of 13 -- 16 keV as well as the background model. There is no hint of a line at 14.4 keV and the expected signal at 90 \% C.L. is shown as the blue line. The model independent limit of $g^{\text{eff}}_{AN}\times g_{Ae}=6.1\times10^{-17}$ is shown in Fig.~\ref{fig:Fig7_Fe57_CB_RD_limit} (b) compared with the EDELWEISS limits \cite{2013_EDELWEISS}. The $g^{\text{eff}}_{AN}$ is model-dependent coupling. In KSVZ models, it dependents on the flavor-singlet axial-vector matrix element $S$ \cite{1997_Adams,1997_Altarelli}. In DFSZ models, besides element $S$ it also dependents on the tan$\beta_{DFSZ}$, which is the ratio of two Higgs vacuum expectation values of the model. The dashed red line in Fig.~\ref{fig:Fig7_Fe57_CB_RD_limit} (a) shows the 90\% C.L. limit at the DFSZ model with $S=0.5$ and cos$\beta^{2}_{DFSZ}=1$. In DFSZ and KSVZ models, using the parameters described above, the axion mass can be constrained to 9 eV/c$^{2}$ and 173 eV/c$^{2}$, respectively.  
Combining the results from CBRD channel and $^{57}$Fe channel, our results exclude the axion mass heavier than 0.9 eV/c$^{2}$ and 173 eV/c$^{2}$ according to the DFSZ and KSVZ model respectively.

\begin{figure}[htb]
  \includegraphics[width=1.0\linewidth]{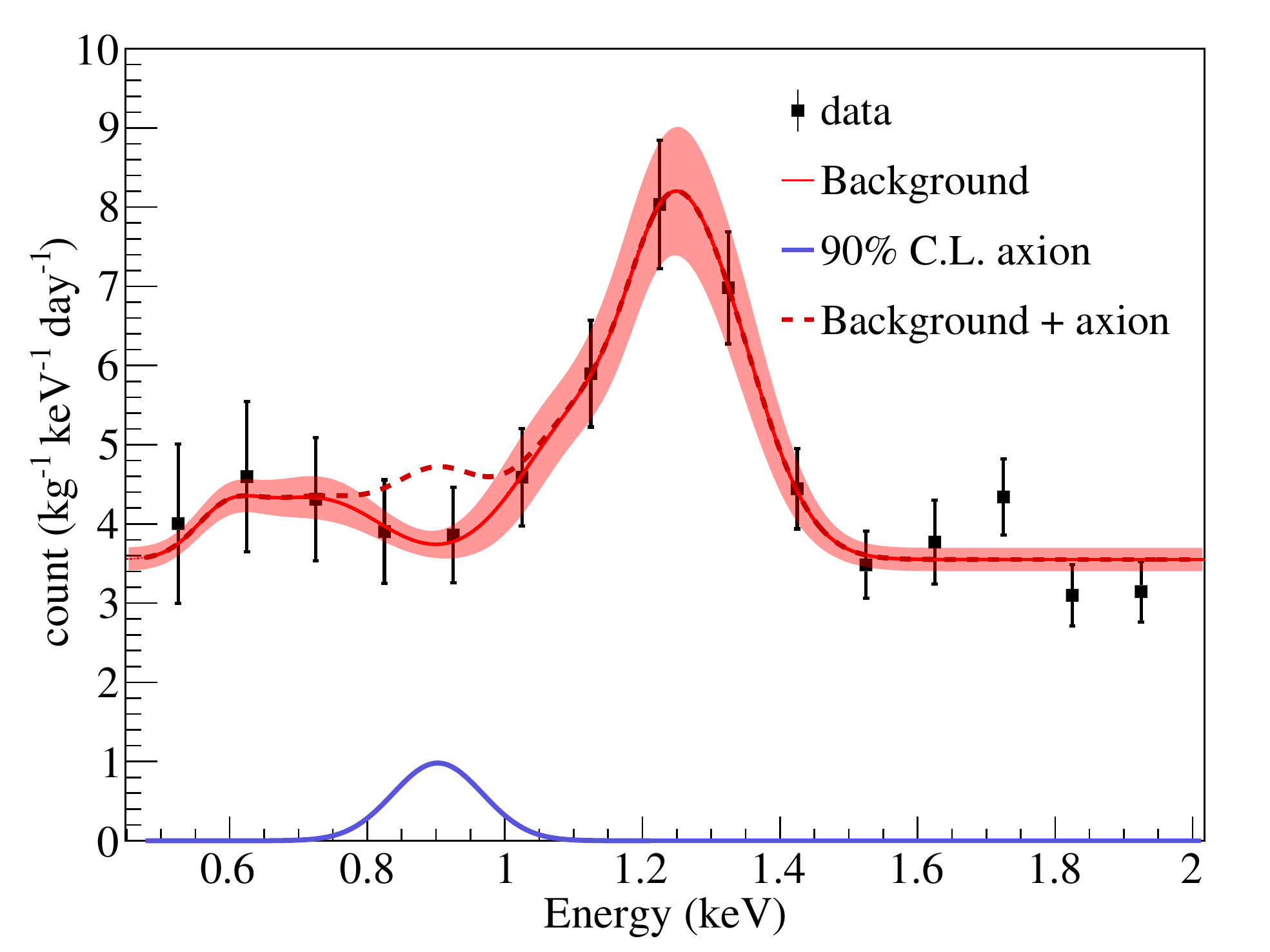}
  \caption{\label{Fig8_DM} The DM ALPs fitting result at the mass of 0.9 keV. Data points: $B_{0}$ bulk spectrum; red line: background assumption: blue line: 90\% C.L. axion signal; dashed red line: 90\% C.L. axion signal superimposes on the background model.}
\end{figure}

\begin{figure}[htb]
  \includegraphics[width=1.0\linewidth]{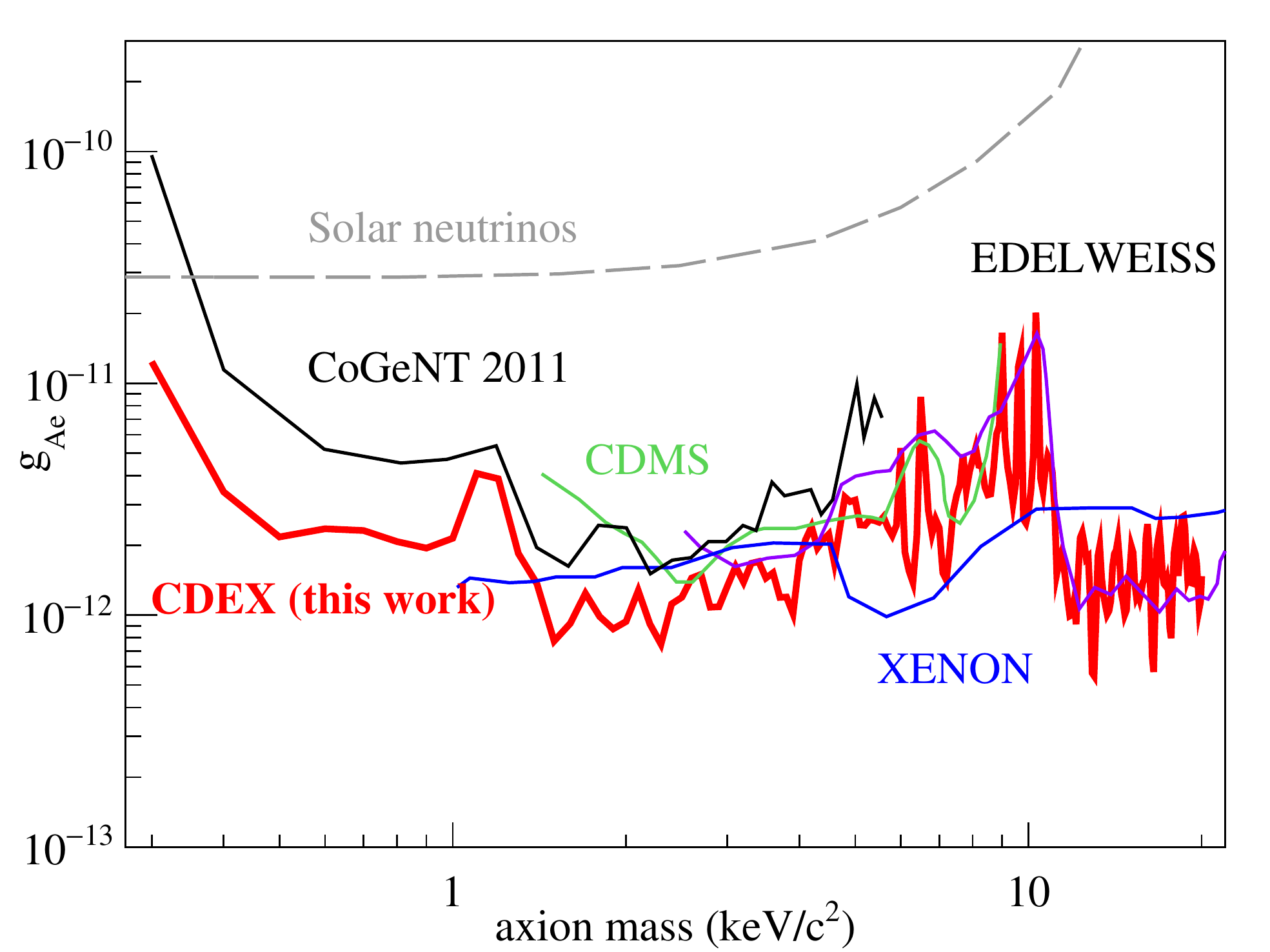}
  \caption{\label{Fig9_DM_EP} The CDEX-1 90\% C.L. on galactic DM axion (red line), together with the bounds from other representative experiments \cite{2011_Aalseth,2009_Ahmed,2014_XENON100,2013_EDELWEISS} and solar neutrinos bounds \cite{2009_Gondolo},}
\end{figure}

\subsection{Galactic ALPs}

No statistically significant excess signals are observed, scanning the energy range between 0.475 -- 20 keV with the same method as $^{57}$Fe solar axion. Fig.~\ref{Fig8_DM} shows one of the fitting results at the mass of 0.9 keV. The 90\% C.L. limit on $g_{Ae}$ is displayed in Fig.~\ref{Fig9_DM_EP}. The peaks in the limit plots corresponds to the K/L x-ray peaks in the spectrum and the steps around 1.3 keV and 10 keV due to the atomic energy levels. Because of the monochromatic signal and the good energy resolution, the limit is sensitive to the fluctuations of  individual bins . 
The CDEX-1 limits are more stringent than improve over from CoGeNT \cite{2011_Aalseth} at axion mass less than 1 keV, due to improved detector threshold, energy resolution and residual background.

\section{Summary AND Prospects}
The CDEX-1 hardware, interested axion sources, the details of data analysis procedures and results have been described.
The limits of $g_{Ae}$ couplings of solar axions and galactic dark matter ALPs are derived with a data size of 335.6 kg-days. We demonstrated that the PCGe detector is a potential technique to axion searches due to the excellent energy resolution and low energy threshold, especially for the peak searches such as the DM ALPs and $^{57}$Fe solar axions. The constraints on DM ALPs below the mass of 1 keV/c$^{2}$ improves over the previous results. 

The CDEX-1 data spanning over 17 months allows the studies of annual modulation. In additional, research efforts on lowering the detector threshold, controlling of radiopurity as well as understanding of background are being pursued. Improved sensitivities of the studies of WIMP dark matter and axions can be foreseen.





\begin{acknowledgments}
  This work was supported by the National Natural Science Foundation of China (Contracts No. 11505101, 11175099, 11275107, 11475117 and 11475099), and the National Basic Research Program of China (973 Program) (2010CB833006), and China Postdoctoral Science Foundation, and MOST 104-2112-M-001-038-MY3, and the Academia Sinica Principle Investigator Award 2011-2015 from Taiwan.
\end{acknowledgments}


\newpage
\bibliography{C1A_axion}

\end{document}